\documentclass[aps,prb,twocolumn, groupedaddress,showpacs]{revtex4}
\usepackage{graphicx,amsmath}
\usepackage{amssymb}

\begin{document}
\title{Mott transition in the Hubbard model on the hyper-kagome lattice}

\author{Hunpyo Lee}
\author{Hartmut Monien}
\affiliation{Bethe Center for Theoretical Physics, Universit\"at Bonn, 53115 Bonn, Germany}
\date{\today}

\begin{abstract}
  Motivated by recent experiment on the Na$_4$Ir$_3$O$_8$ compound 
  we study the Hubbard model on the ``hyper-kagome lattice'', which forms a
  three-dimensional network of corner sharing triangles, using dynamical
  cluster approximation (DCA) method with $N_c$=12 combined with the continuous-time
  quantum Monte Carlo (CT QMC) method. The system undergoes a Mott
  transition if the Hubbard interaction $U/W$ ($W$ is the bandwidth) exceeds the value of 1.2 for $T$=0.1 and displays
  reentrant behavior due to competition between the magnetic correlation
  and the kinetic energy of electrons due to the geometrical frustration. We
  observe a ``critical slowing down'' of the double occupancy
  which shows evidence of a continuous transition. The nearest-neighbor and next nearest-neighbor
  spin-spin correlations indicate a paramagnetic metallic state in the
  weak-coupling regime and an antiferromagnetic (AF) Mott insulator in the
  strong-coupling regime within the temperature range
  which we can access with our numerical tools. 
\end{abstract}
 
\maketitle

Frustrated electronic systems exhibit a wide range of phases due to
competition between strong electronic correlations and geometric
frustration. The Na$_4$Ir$_3$O$_8$ compound with a S=1/2
three-dimensional network of corner sharing triangles, which is called the
``hyper-kagome lattice'', was discovered.\cite{Okamoto07} Due to the
structure of corner sharing triangles, the system displays the electronic
correlations and geometric frustration at the same time. Each tetrahedron in
the pyrochlore lattice, which is shown in Fig. 1, consists of Ir (indicated as
``filled'' circles) and Na (indicated as ``empty'' circles). S=1/2 spins are
carried through Ir$^{4+}$ on the hyper-kagome lattice (blue lines in Fig. 1) because Ir is
tetravalent with five electrons in $5d$ orbitals. In recent experiment by
Y. Okamoto (Ref. 1), a spin liquid state at low temperature was
found. Theoretical work, related to this experiment, has mostly addressed the
strong-coupling limit using the Heisenberg model\cite{Balents08,Chen08,Lawler08,Hopkinson07,Zhou08}. These
papers concentrated on the nature of the ground state which was identified as
the spin liquid state. Even more recently, it was shown experimentally that this material
undergoes a transition to a metallic state under pressure\cite{Takagi08}. 

The two important theoretical questions which we will address in this Letter are
if the phase diagram shows a reentrant Mott phase with a first-order
transition as was found for the anisotropic triangular lattice \cite{Ohashi08} and if the spin
liquid state is stable at finite temperatures. The dynamical mean field theory (DMFT) \cite{Metzner89,Georges96} is a successful
method to describe the Mott transition. However, since the single-site DMFT ignores the
spatial fluctuations, the system becomes metallic as temperature is decreased. This behavior
is quite different from the phase diagram obtained from quantum Monte Carlo (QMC)
on the triangular lattice.\cite{Bulut05,Aryanpour06} The cluster extension of
the DMFT methods such as the dynamical cluster approximation (DCA)
\cite{Maier05,Hettler98,Hettler00} and 
the cellular dynamical mean field theory (CDMFT) \cite{Kotliar01} were developed in order to
consider the spatial fluctuations. In these methods
the short range correlations are treated exactly within the cluster size, while
the long range correlations are included on the mean field level only. The
difference between both methods lies in the boundary condition of the cluster
given by the Laue function \cite{Maier05}. In this paper, we employ the DCA
method combined with the continuous-time quantum Monte Carlo (CT QMC) method \cite{Rubtsov05} as
cluster solver and measure the total density of states (DOS), double occupancy, nearest-neighbor
and next nearest-neighbor spin-spin correlation functions. 

\begin{figure}
\includegraphics[width=150pt]{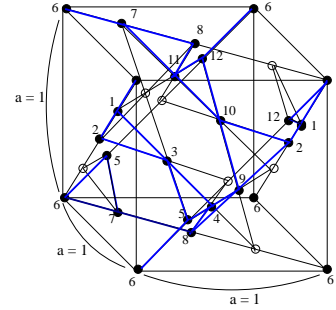}
\caption {In the pyrochlore lattice, ``filled'' circles and ``empty'' circles
  present Ir and Na, respectively. The blue lines show the hyper-kagome
  lattice of Ir. a is the lattice constant. Lattice points labeled by 1,2,...,12 represent
relative positions in each unit cell.}
\end{figure} 

We consider the Hubbard model on the hyper-kagome lattice at half-filling: 
\begin{equation}
  H=-t\sum_{\langle i,j \rangle \sigma}c^{\dag}_{i\sigma}c_{j\sigma} 
+ U \sum_{i}n_{i\uparrow}n_{i\downarrow},
\end{equation} 
where $c_{i\sigma}(c^{\dag}_{i\sigma})$ is the annihilation (creation) operator
of an electron with spin $\sigma$ at the $i$-th site, $t$ is the hopping matrix
element and $U$ represents the Coulomb repulsion. We set $t$=1.0 and the
bandwidth is $W$=$6t$. For the DCA method we consider
an unit cell which has twelve sites ($N_c$=12) in the simple cubic lattice, as shown in
Fig. 1. In the DCA method the Green's functions are determined by the self-consistency
condition. We directly write the hopping matrix $T(\tilde{{\bf k}})$ with the
periodic boundary condition in the real space. Here
$G_{0}$, $\Sigma$ and $T(\tilde{{\bf k}})$ are described by 12 $\times$ 12 matrices. In this
case the self-consistent equation for the DCA method is the same as for the CDMFT
method except the hopping matrix $T(\tilde{{\bf k}})$:
\begin{eqnarray} \label{self-e}
G_{0}^{-1}(i\omega_{n}) &=&\left(\int d\tilde{{\bf k}}%
\frac{1}{i\omega_{n}+\mu -T(\tilde{{\bf k}})-\Sigma (i\omega _{n})}\right) ^{-1}%
\kern-1em  \notag \\
&+&\Sigma (i\omega _{n}),
\end{eqnarray}
where $\mu$ is the chemical potential and $\omega_{n}$ is the Matsubara
frequency. The Brillouin zone of the superlattice also becomes simple cubic
lattice with $\pi$/a $<$ $k_x$, $k_y$, $k_z$ $<$ $\pi$/a, where a = 1 is the
lattice constant and the summation of $\tilde{{\bf k}}$ is taken over the
Brillouin zone of the superlattice. We use 8 $\times$ $10^{6}$ $\tilde{\bf
  k}$-points for integration of Eq. (\ref{self-e}). We solve the quantum
impurity problem using the CT QMC method. Unlike the Hirsch-Fye discrete time
quantum Monte Carlo method \cite{Hirsch86} which is carried out by the
auxiliary Ising-like spins, the CT QMC method performs a random walk in the
space of the perturbation expansion without Trotter decomposition error.
The computational time of CT QMC scales as $\langle k\rangle \thicksim
0.5N_cU\beta$ and is therefore superior to the Hirsch-Fye quantum Monte Carlo method
for which the computational time scales as $\langle$k$\rangle$ $\thicksim$ 2$N_cU\beta$, where
$\langle k\rangle$ is the matrix size.  We briefly summarize the CT QMC
method. The main idea of CT QMC is to perform a diagram expansion of the
partition function $Z=Z_{0} e^{-U\int d\tau n_{\uparrow}n_{\downarrow}}$ in
powers of the interaction $U$. In this case we can reexpress the partition
function as
  \begin{equation}
    \begin{aligned}
      {\cal Z} = \sum_{k}Z_{0}\frac{(-U)^{k}}{k!}\int d\tau_{1}\cdots d\tau_{k}
      \int{\cal D}[c,\bar{c}]\\
      {<{n_{\uparrow}(\tau_{1})n_{\downarrow}(\tau_{1})\cdots
        n_{\uparrow}(\tau_{k})n_{\downarrow}(\tau_{k})}>},
    \end{aligned}
  \end{equation}
where ${<{n_{\uparrow}(\tau_{1})n_{\downarrow}(\tau_{1})\cdots
n_{\uparrow}(\tau_{k})n_{\downarrow}(\tau_{k})}>}$ is determined by the
non-interacting Green's function and Wick's theorem and $Z_{0}$ is the
unperturbed part. The interacting Green's functions are
calculated by numerical averaging of Eq. (3). We use $3 \times
10^{6}$ QMC sweeps approximately and consider twenty-one Matsubara frequency points. 
\begin{figure}
\includegraphics[width=220pt]{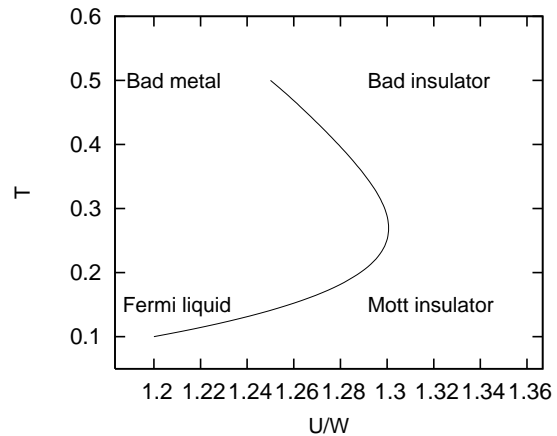}
\caption{Phase diagram of the Hubbard model on the hyper-kagome lattice}
\end{figure}

Now let us investigate the phase diagram of the hyper-kagome lattice Hubbard
model at half filling. In order to check the existence of a gap, we calculate
the density of state (DOS) at the Fermi level ($\omega$ = 0) by using 
the Pade approximation and the analytic form which is given by
\begin{equation}
G(\beta/2) \approx \frac{1}{\beta}\rho(\omega=0),
\end{equation}
where $\beta$ is the inverse temperature and $\rho(\omega)$ is the density of
states. Because Eq. (4) is satisfied at
sufficiently low temperature, we employ the Pade approximation in $T$$>$0.1
regions and compare the results of both methods at $T$=0.1. The Pade
approximation method for analytical continuation works well at the high
temperature regime
due to very small error in CT QMC method. Moreover, the results of Eq. (4) support 
those obtained from the Pade approximation at low temperature. The phase
diagram in U-T plane is shown in Fig. 2. As temperature is decreased in the
transition region ($U/W$=1.28) above $T$=0.3, the system prefers the
metallic state due to the dominant low energy excitation of electrons. Below $T$=0.2
the system is strongly controlled not by itinerant electrons but by
antiferromagnetic (AF) fluctuations, which are ignored in the single-site DMFT. Therefore, the Mott insulator lies in the low
temperature regions. This results in a reentrant behavior which is also presented in the
Hubbard model on the anisotropic triangular lattice \cite{Ohashi08} related to the frustrated organic material
$\kappa$-(BEDT-TTF)$\mathrm{_2Cu[N(CN)_2]Cl}$ \cite{Lefebvre00,Kagawa04}.
\begin{figure}
\includegraphics[width=250pt]{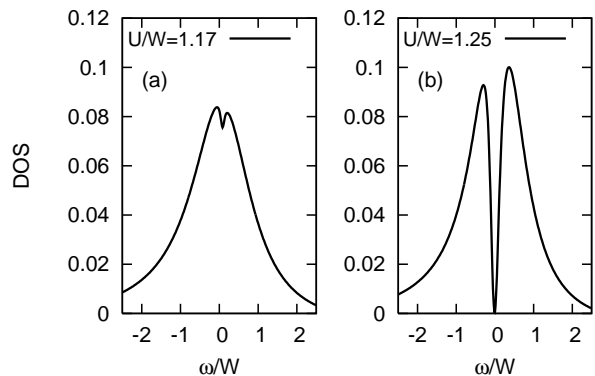}
\caption{The density of state corresponding to (a) $U/W$=1.17 and (b) $U/W$=1.25 for
  $T$=0.1. The Pade approximation is employed for analytical continuation}
\end{figure}
We analyze the DOS close to critical $U/W$ at $T$=0.1 using Pade approximation for
analytical continuation. The DOS for $U/W$=1.17 and $U/W$=1.25 is exhibited in
Figs. 3(a)-(b). At $U/W$=1.17 a quasiparticle peak due
to itinerant of electrons is formed strongly
around the Fermi level. At $U/W$=1.25 the DOS changes dramatically at the Fermi
level and the system turns to a Mott insulator. We did not observe any evidence for the
pseudogap formation like the results of kagome lattice \cite{Ohashi06}. The
magnetic state is suppressed due to the geometrical frustration in contrast
to the square lattice \cite{Moukouri01}. To investigate the character of Mott
transition in more detail we
present the double occupancy as a function of temperature for various
interactions in Figs. 4(a)-(c). As temperature is decreased at weak interaction strength $U/W$=1.0
in Fig. 4(a), the double occupancy is increased. This means
that the entropy due to kinetic energy of the dominant quasiparticles, is optimized by pushing the
interaction in order to decrease the free energy. Non-monotonic dependence of the double occupancy is observed at
intermediate interaction $U/W$=1.2 in Fig. 4(b).
\begin{figure}
\includegraphics[width=190pt]{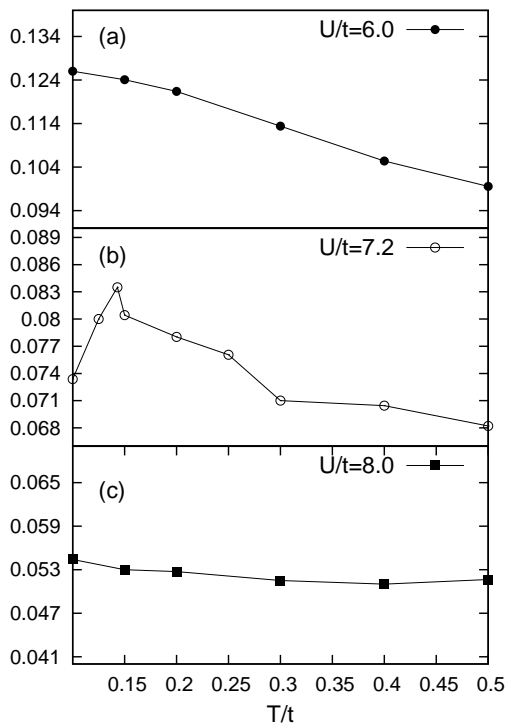}
\caption{Double occupancy as a function $T$ for (a) $U/W$=1.0, (b) $U/W$=1.2 and
  (c) $U/W$=1.333.}
\end{figure}
\begin{figure}
\includegraphics[width=220pt]{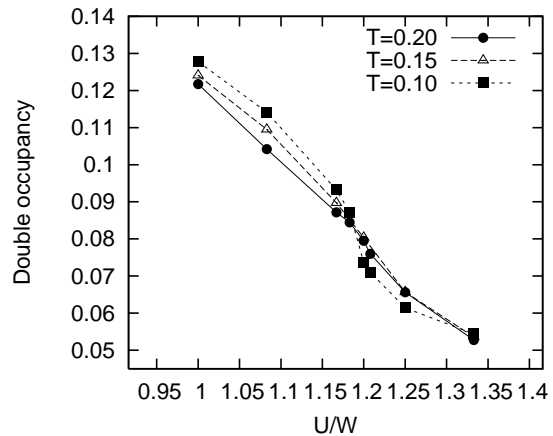}
\caption{Double occupancy as a function $U/W$ for $T$=0.2 and $T$=0.1. There is
  a ``critical slowing down'' behavior at $T$=0.1 which represents the evidence of
  a continuous transition. $U_c/W$=1.2 for $T$=0.1}
\end{figure}
For $T>$0.3 the system displays the
insulating state due to effect of the dominant local magnetic moment. As temperature is
decreased, it changes into the metallic state due to development of the
quasiparticle. After passing the peak
temperature $T_0$=0.15, at lower temperatures the system reenters the
insulating state because the magnetic correlations get enhanced. This is indeed
evidence for the reentrant behavior which is exhibited in geometrically
frustrated systems like the case of Hubbard model on anisotropic triangular
lattice. In Fig. 4(c), we see a small gradual
increase in the double occupancy as temperature is decreased, reflecting a
gain in kinetic energy. Our analysis in the Fermi liquid region
($U/W$=1.0) and Mott insulator region ($U/W$=1.33) is similar to that obtained
from the cluster-extension of DMFT method on the square lattice \cite{Gull08}. We further investigate the double
occupancy as function of interaction at various temperatures in order to see the type of
transition. The result is shown in Fig. 5. As interaction is increased, at $T$=0.2 the double occupancy
decreases smoothly because of the development of local magnetic moment. At $T$=0.1 the ``critical slowing down'' behavior
is observed around $U_c/W$=1.2. This behavior is quite similar to the results of the
fully frustrated Hubbard model which indicates a second-order transition in the
single-site DMFT calculation \cite{Rozenberg99}. We suspect that there is a continuous transition in the Hubbard model on the
hyper-kagome lattice. We calculate the nearest-neighbor $\langle S^z_i S^z_{i+1} \rangle$ and next
nearest-neighbor spin-spin correlation $\langle S^z_i S^z_{i+2} \rangle$, which are
shown in Fig 6. In the weak interaction regions the system displays
paramagnetic metallic behavior, unlike the case of the square lattice which has
strong AF correlations due to perfect nesting. As interaction is increased, there is a competition between
the quasiparticle formation and the frustrated spin correlation. At $U_c/W$=1.27
for $T$=0.2 both the nearest-neighbor and next nearest-neighbor spin-spin correlations are enhanced
at the same time. We find the AF Mott insulator in the strong interaction
regions. We do not find a spin liquid state for $T$$>$0.1. 

Finally, we compare our results with those recently obtained from
the mean-field calculation\cite{Podolsky08}. In the mean-field calculation the
system has continuous metal-insulator transition at critical interaction
$U_c/W$=1.03. These results are comparable to our results with a
continuous metal-insulator transition at $U_c/W$=1.2 for $T$=0.1.  
\begin{figure}
\includegraphics[width=240pt]{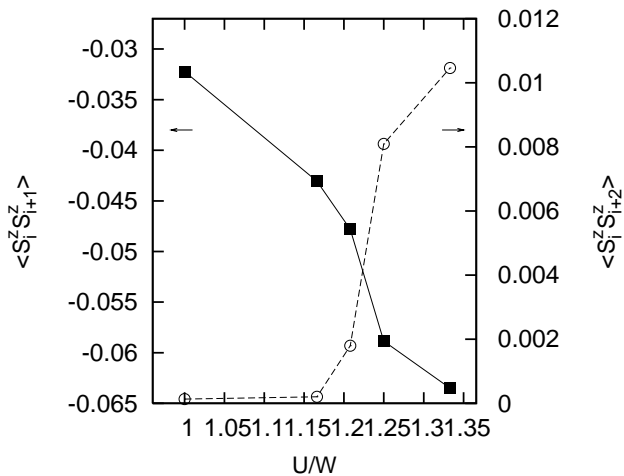}
\caption{The nearest-neighbor and next nearest-neighbor spin-spin correlation function as a function of $U/W$
  for $T$=0.2. Both spin-spin correlations are rapidly increased at $U_c/W$=1.27}
\end{figure}

In summary, we have presented results of DCA study combined with CT QMC method as
a cluster solver in the Hubbard model on the hyper-kagome lattice. The phase
diagram in the U-T plane has reentrant behavior due to geometrical
frustration and is quite similar to that on the anisotropic triangular lattice. The
difference of both lattice types is that while the hyper-kagome lattice exhibits a
continuous transition, the anisotropic triangular lattice
displays a first-order transition. Using Pade approximation for the analytical
continuation we calculated the DOS close to critical $U_c/W$=1.2 for $T$=0.1. There is
rapid change in the DOS at Fermi level. The formation of pseudogap is not
observed because the magnetic fluctuations are suppressed. We investigated the
double occupancy as a function of temperature and as a function of
interaction $U/W$ to study the nature of the Mott transition in more detail. For
intermediate values of the interaction, the double occupancy as a function of temperature has
non-monotonic behavior which represents the character of reentrant behavior. We also
find the ``critical slowing down'' of the double occupancy as a function of
interaction. We suspect that at low temperature there might be a
signal of a continuous transition. Finally, we calculated the nearest-neighbor
and next nearest-neighbor spin-spin correlations as a function of
interaction. We observed a paramagnetic metallic state in the
weak-coupling regime. When the interaction passes the critical point, both nearest-neighbor and next
nearest-neighbor spin-spin correlations are rapidly increasing. In the strong-coupling regions the system shows the AF Mott
insulator state. We did not observe a spin liquid state (paramagnetic Mott
insulator) at lower temperature. However, from our calculation we expect
that a continuous Mott transition with reentrant behavior will be observed experimentally.            

We would like to thank A. J. Millis for helpful discussions.
\bibliography{paper}

\end{document}